\documentclass{aa}  
\usepackage{graphicx,epsf,epsfig,float}
\usepackage{txfonts}
\def\ion#1#2{{\rm #1}~{\sc #2}}

\def\nom{Gon{\c{c}}alves}
\def\Agata{R\' o\. za\' nska~}
\def\los{line-of-sight}

\def\eg{{e.g.}}
\def\ie{{i.e.}}
\def\vs{{vs.}}
\def\kms{km\,s$^{-1}$}

\def\ergs{erg\,s$^{-1}$} 
\def\chandra{{\it Chandra}}
\def\xmm{{\it XMM-Newton}}
\def\xi{$\xiup$}
\def\NH{$N_{\rm H}$}
\def\nh{$n_{\rm H}$}
\def\xiun{${\rm erg\,cm\,s^{-1}}$}
\def\NHun{${\rm {cm}^{-2}}$}
\def\nhun{${\rm {cm}^{-3}}$}

\def\vturb{$v_{\rm turb}$}
\def\vflow{$v_{\rm flow}$}
\def\vtherm{$v_{\rm th}$}

\begin{document}

   \title{A new model for the Warm Absorber in NGC 3783: \\ 
          a single medium in total pressure equilibrium}


   \author{A.\/C. \nom\
          \inst{1,2}
          \and
          S. Collin\inst{1}
	  \and
	  A.-M. Dumont\inst{1}
	  \and	 
	  M. Mouchet\inst{1,3}
	  \and 
          A. R\'o\.za\'nska\inst{4}
	  \and
          L. Chevallier\inst{1,4}
 	  \and
	  R.\/W. Goosmann\inst{1}
	  }

   \offprints{A.\,C. \nom}

   \institute{
LUTH, Observatoire de  Paris-Meudon, 
5 Place Jules Janssen, 92195 Meudon Cedex, France  
(\email{anabela.goncalves@obspm.fr}) 
         \and
CAAUL, Observat\'orio Astron\'omico de Lisboa,
Tapada da Ajuda, 1349-018 Lisboa, Portugal
         \and
APC, Universit\'e Denis Diderot 
Paris 7, 75005 Paris, France
         \and 
Nicolaus Copernicus Astronomical Center,
Bartycka 18, 00-716 Warszawa, Poland
          }
   \date{Received January 12, 2006; accepted Mars 14, 2006}

   \abstract
    {
Many active galactic nuclei exhibit X-ray features 
typical of the highly ionized gas called ``Warm Absorber'' 
(WA). Such a material appears to be stratified, displaying zones of 
different density, temperature, and ionization. In this paper,   
we investigate the possibility of modelling the WA gas in 
NGC~3783 as a single medium in total pressure equilibrium.}  
   {Our goal is to demonstrate that the WA can be well 
modelled assuming constant total pressure, in contrast to  
the current descriptions that are based on the presence 
of multiple regions, each in constant density. The assumption 
of total pressure equilibrium yields a more physical description  
of the WA, resulting in the natural stratification 
of the ionized gas, and providing an explanation for the 
presence of lines from different ionization states, as 
observed in WA spectra.}
   {We have used the photoionization code TITAN, developed 
by our team, to compute a grid of constant total pressure 
models with the purpose of fitting the WA in NGC~3783. 
We have compared 
our models to the 900 ks \chandra\ spectrum of NGC~3783 
and to previous studies where the WA was described by  
multiple zones of constant density.}
   {In the case of NGC 3783, the WA features can 
be well reproduced by a clumpy, ionized gas with  
cosmic abondances, ionization parameter $\xiup = 
2500~\mathrm{erg\,cm\,s^{-1}}$, column density 
${N_{\rm H}} = 4\,10^{22}~\mathrm{cm^{-2}}$, 
and constant total pressure.}
   {We have shown that the WA in NGC 3783 can be  
modelled by a single medium in total pressure equilibrium; 
this is probably the case for other 
WAs currently described by multi-zone, 
constant density models. In addition, our work 
demonstrates that the TITAN code is well adapted to 
the study of the WA in active galactic nuclei, 
opening new prospects for the use of TITAN  
by a larger community. }

   \maketitle


\section{Introduction}

Many Active Galactic Nuclei (AGN) exhibit important 
 X-ray absorption features caused by the presence of highly 
ionized gas located in the \los\  of the central continuum; 
this material is called ``Warm Absorber'' (hereafter WA). 
The first observations of WA gas in AGN were reported by 
Halpern et al. (1984) in the {\it Einstein Observatory} 
spectrum of MR~2251$-$178, a quasar displaying a large 
absorption feature around 1~keV; such a feature is consistent 
with the presence of gas photoionized by the hard X-rays 
produced near the AGN central engine.

Early {\it ASCA} observations have revealed 
the presence of  intrinsic absorption   
in $\sim$\,50\%\ of type~1 Seyferts. With the advent 
of space X-ray observatories such as \chandra\  and 
\xmm, which carry aboard high-resolution grating 
spectrographs, an important set of high quality 
data became available; evidence for the 
presence of  warm gas was found in many other types 
of AGN, \eg\  Seyfert~2s (Kinkhabwala et al. 2002), 
Narrow Line Seyfert 1s 
(\eg\  Matsumoto et al. 2004), BAL~QSOs 
(Guainazzi et al. 2001; Grupe et al. 2003), and 
even some BL~Lacs (\eg\  Sambruna et al. 1997), 
although opinions diverge for these 
objects (Blustin 2004). In Seyfert 1s, the WA spectra 
revealed the presence of tens of absorption lines, 
covering a wide range of ionization states and 
blueshifted by a few hundreds to thousands of \kms\ 
(an indication that the absorbing material is 
outflowing). In type~2 AGN, the data have shown 
the presence of emission lines.

Despite the undeniable improvements in our knowledge of 
the WA, some important issues remain a subject of debate, 
namely: {\it (i)} the geometry (multi-zone gas? 
clumps or flow?) and location of the 
WA, {\it (ii)} the physical conditions of the 
absorbing/emitting gas, and {\it (iii)} the importance 
of the WA in the energetics of the AGN. 
Trying to solve these questions requires not only 
high quality observations, but also an adequate 
modelling of the X-ray data through the use of 
reliable photoionization codes, which calculate the full 
radiative transfer and allow for a good physical 
description of the ionized gas. 

The WA gas appears to be 
stratified, displaying multiple zones of different 
density, temperature, and ionization  (\eg\  Krolik \& 
Kriss 2001; R\'o\.za\'nska et al. 2006).  Until now, 
the WA in AGN has been modelled assuming the presence 
of several regions,  each region being modelled in 
constant density and having a different 
ionization parameter\footnote{The ionization parameter 
is defined as $L/n_{\rm H}R^{2}$, where $L$ 
is the source's bolometric luminosity, 
$n_{\rm H}$ is the hydrogen number density at the illuminated 
face of the medium, and $R$ is the distance from the WA 
to the illuminating source.} ($\xiup$, in units 
of \xiun), column density (\NH, in units of \NHun), 
etc. However, a stratified medium can be obtained 
naturally, if one assumes the gas to be in pressure 
equilibrium. Such a medium could explain 
the presence of lines from different ionization 
states and account naturally for the other properties 
of a model composite of several constant density regions. 

We have addressed the above-mentioned  points through 
the study of the WA in NGC~3783, for which unrivaled 
quality \chandra/HETGS data (Kaspi et al. 2002) 
are available in the archives. 
There is a short discussion of the WA in NGC~3783 in 
the next section. We have modelled the data using the 
photoionization code TITAN, briefly described in Sect.~3.  
Our first results are given in Sect.~4, and the 
conclusions in Sect.~5.

\section{The Warm Absorber in NGC~3783}

NGC~3783 is a bright ($V$\,$\sim$\,13.5), nearby 
($z$\,=\,0.0097), Seyfert~1.5 galaxy observed in the 
Optical, UV and X-rays. The WA in this object has been 
discussed by several authors (\eg\  Kaspi et al.~2001,\,2002; 
Netzer et al.~2003; Krongold et~al.~2003; Behar et al.~2003) 
based on \chandra\ and \xmm\ data. These 
studies seem to agree on the presence of a multi-zone 
gas constituted by three or more regions of different ionizations.  
In terms of the kinematics of the WA, two or more velocity 
systems have been identified in \chandra\ observations; 
they are compatible with those observed in UV 
spectra. A  single outflow velocity system 
(\vflow\,$\sim$\,600--800~\kms) 
seems enough to describe 
\xmm\ observations, which have lower spectral resolution. 
There is no consensus on  
a possible correlation between the velocity shifts, or the 
$FWHM$s, with the ionization potentials of the ions.

Although the WA in NGC~3783 has been the object of many 
studies, these have assumed either a constant density  
(\eg\  Netzer et al. 2003), or a dynamical state (Chelouche 
\& Netzer 2005) for the modelling. In addition, they all 
require multiple zones of different density, temperature,  
and ionization, which are invoked to explain the large 
span in ionization observed in the WA spectrum.  
Furthermore, when plotted on the S-curve of thermal equilibrium 
$log\,(T)$ \vs\ $log\,(\xiup/T\,)$ (where $T$ is the 
temperature of the medium and $\xiup$ is the ionization parameter),  
these zones lie on a vertical line of roughly the same gas 
pressure, suggesting that a single medium in pressure 
equilibrium could account for the WA in this object. 
A physical modelling of the gas through adequate photoionization 
codes remains necessary to confirm or to invalidate 
such an assumption, that being the main goal of this paper. 

We note that, for relatively thin slabs, the 
radiation and gas pressure remain constant, the 
temperature does not vary much within the slab, 
and constant pressure models can thus be 
approximated by a constant density model. 
However, for thicker media, the ratio between the 
gas and the radiation pressure varies across the 
slab, inducing variations in the density and the 
temperature. In such cases, it is important to take 
the total (gas+radiation) pressure into account.

\section{Modelling the WA with the TITAN code}

We have used the photoionization code TITAN, 
developed by our team (Dumont et al. 2000,\,2002; 
Collin et al. 2004), to model the 900 ks \chandra/HETGS  
observations and to constrain the physical conditions 
of the WA gas in NGC~3783. 

TITAN is well-suited both for the study of optically thick 
(Thomson thickness up to several tens) and thin media, 
such as the WA. It computes the gas structure in thermal 
and ionization equilibria, giving as output the 
ionization and temperature structures, as well as the 
reflected, emitted outward, and absorbed spectra. 
One advantage of TITAN 
is that it treats the transfer of both the lines 
and the continuum using the ALI  (Accelerated 
Lambda Iteration) method, which 
very precisely computes line and continuum fluxes.  
Our line transfer treatment leads to a 
longer computation time than with other approximate 
methods; we thus use fewer lines than those available 
in other codes. 
Our atomic data include $\sim$\,$10^{3}$ lines from ions 
and atoms of H, He,  C, N, O,  Ne,  Mg, Si, S, and Fe; 
they omit features like the Unresolved 
Transition Array (UTA) and the inner shell transitions 
(except the iron K lines). In the case where the radiation 
pressure and the losses from the lines are only a 
few tens percent of those from the continuum, as is the  
case in this study, the missing lines do not change the ionization 
and thermal structure much (Chevallier et al. 2006).  

%

The total (gas\,+\,radiation) 
pressure is computed after the plasma temperature has been 
determined by the thermal balance equation. The pressure is kept 
fixed within the medium, thus determining the density profile. 
The model is iterated until convergence 
(see Dumont et al. 2000,\,2002). 
In the presence of an external, strong, irradiating source, such as 
the hard X-rays ionizing the WA material, the irradiated medium 
develops an ionization instability (Krolik et al. 1981). If the 
medium is in pressure equilibrium, it leads to a temperature 
discontinuity (\eg\  Ko \& Kallman 1994) with at least  
three states of thermal equilibria (high-, intermediate-, 
and low-temperature), the middle one being unstable. 
Our approach to the thermal instability problem consists 
of adopting an equilibrium temperature value between the 
``hot'' and the ``cold'' stable solutions. Such a method 
provides a unique solution 
(for a more complete discussion of the thermal instabilities 
with TITAN, see Sect.~4.1.3 in \Agata\ et al. 2002).

Our models assume a 1D plane-parallel geometry, with 
the WA medium being illuminated on one side by an incident 
continuum. We assume the incident radiation field to 
be concentrated in a very small pencil-like shape 
centered on the normal direction. 
The models are parameterized by 
the ionization parameter $\xiup$, the column density \NH, 
the hydrogen number density \nh, and the incident flux Spectral 
Energy Distribution (SED). We can also consider a non-zero 
turbulent velocity \vturb. 

Concerning  NGC~3783 specifically, we note first that the 
turbulent velocity deduced from the observations 
(\eg\  Kaspi et al. 2000) is highly supersonic 
compared to the thermal velocity (\vtherm) in the absorbing 
medium. It is possible to address this situation either  
$(i)$ by taking into account the strong dissipation occuring through 
shock waves in the supersonic medium (this leads to a higher 
temperature than in a purely photoionized case),  or 
 $(ii)$ by assuming a photoionized, inhomogeneous medium with several 
clumps lying in the \los. These gas clumps have an average 
relative velocity, that we identify as \vturb.     
We thus assume the WA in NGC~3783 to consist of such 
gas blobs embedded in a surrounding, hotter medium.  
The clumpy WA medium is assumed to be in total pressure equilibrium.  
We include the gas and the radiation pressure in the total 
pressure, but not the turbulent pressure resulting 
from taking a non-zero \vturb, since it is a quantity exterior 
to the clumps.  
One should, however, take the effects of \vturb\ into account 
when computing the radiative transfer; this has been done for 
all our models. 

The turbulent velocity acting on the line transfer is typically 
one order of magnitude larger than 
the thermal velocity of the heavy elements.  
The optical depth being inversely proportional to \vturb,   
this difference in velocity can be accounted for if each 
clump has a column density $\le$\,\NH/$n$ (where \NH\ 
corresponds to the column density of the clumpy system 
and $n$ is the number of clumps in the line-of-sight). 
From the observer's point of view, and for the 
purposes of our modelling, these clumps can 
be treated just like a continuous medium with column 
density \NH, as each clump receives the 
radiation transmitted - and returned -  by the other ones, 
while the hot surrounding medium does not have any effect on 
the radiation transfer. 

We have computed a grid of 16 constant total pressure models 
to fit the WA in NGC~3783, covering the combinations 
of the 4 possible values of the ionization parameter 
($2000 < \xiup < 3500$) and the total column density 
($3\,10^{22} < N_{\rm H} < 6\,10^{22}$). The \nh\ was 
set to $10^{5}~\rm{cm^{-3}}$ (Netzer et al. 2003) and the 
\vturb\ to 150~\kms (Kaspi et al. 2000). The incident SED 
used, given in Kaspi et al. (2001), covers the 0.2\,eV--400 
keV range. The models were computed using two different 
sets of abundances (Allen 1973; Netzer et al. 2003).

\section{Results and discussion}

Our study shows that the WA can be modelled under total 
pressure equilibrium conditions,  and provides a best model 
with $\xiup$\,=\,2500\,\xiun, $N_{\rm H}$\,=\,4\,10$^{22}$\,\NHun, 
and cosmic (Allen 1973) abundances. While the results are stable  
for a $n_{\rm H}$ up to $10^{12}$~\nhun, the turbulent velocity 
of the clumpy medium seems to play a non-negligible role 
in the modelling. 
Our best model gives a reasonably good fit to the observed data,   
both for the continuum  (following its overall shape up to 
10~keV and reproducing the \ion{O}{vii} and \ion{O}{viii} edges) 
and the lines (both from high and low ionization). 
The observed and modelled spectra, as well as the 
corresponding $EW$s, will be presented in a forthcoming paper 
(\nom\ et al., in preparation), in which the effects of 
using different abundances and \vturb\ values will also be discussed.

\begin{figure}
   \centering
   \vspace{2.9mm}
   \includegraphics[width=6.5cm, angle=-90]{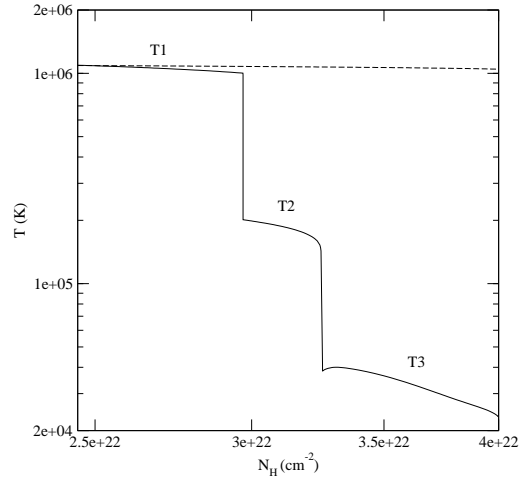}
      \caption{
       The solid line represents the temperature 
       profile resulting from our modelling of the WA as a single 
       medium in total pressure equilibrium. In this case, 
       the temperature discontinuities arise naturally. The dashed line 
       corresponds to a single-zone WA with the same parameters,  
       only modelled under constant density.}   
       \label{Temp}
\vspace{-5.mm}
\end{figure}

For all the models in our grid,  irradiation by the incident 
SED resulted in the formation of a hot surface zone, with an 
almost constant temperature,  followed by a rapid drop by 
one order of magnitude at a certain depth, and a second 
temperature drop deeper in the slab. In Fig.~\ref{Temp} 
we give the temperature profiles for a constant density 
and a constant total pressure model, computed assuming 
the same parameters. We observe that, in the case of a 
constant density model, the temperature remains more 
or less constant along the medium.  Thus, if a composite (\ie\  
multi-component) constant density model is assumed, 
as in Netzer et al. (2003), a relatively flat temperature 
profile is found for each component, with a different 
temperature value for each one.  
When a single medium in constant total pressure is 
assumed, the temperature drops arise naturally; such 
variations in temperature are accompanied by ionization 
fronts, resulting in different ionization fractions for 
the three sections (T1, T2, and T3) visible in the plot.

\begin{figure*}[t]
\vspace{3mm}
\begin{flushleft}
  \resizebox{13.2cm}{!}{\includegraphics{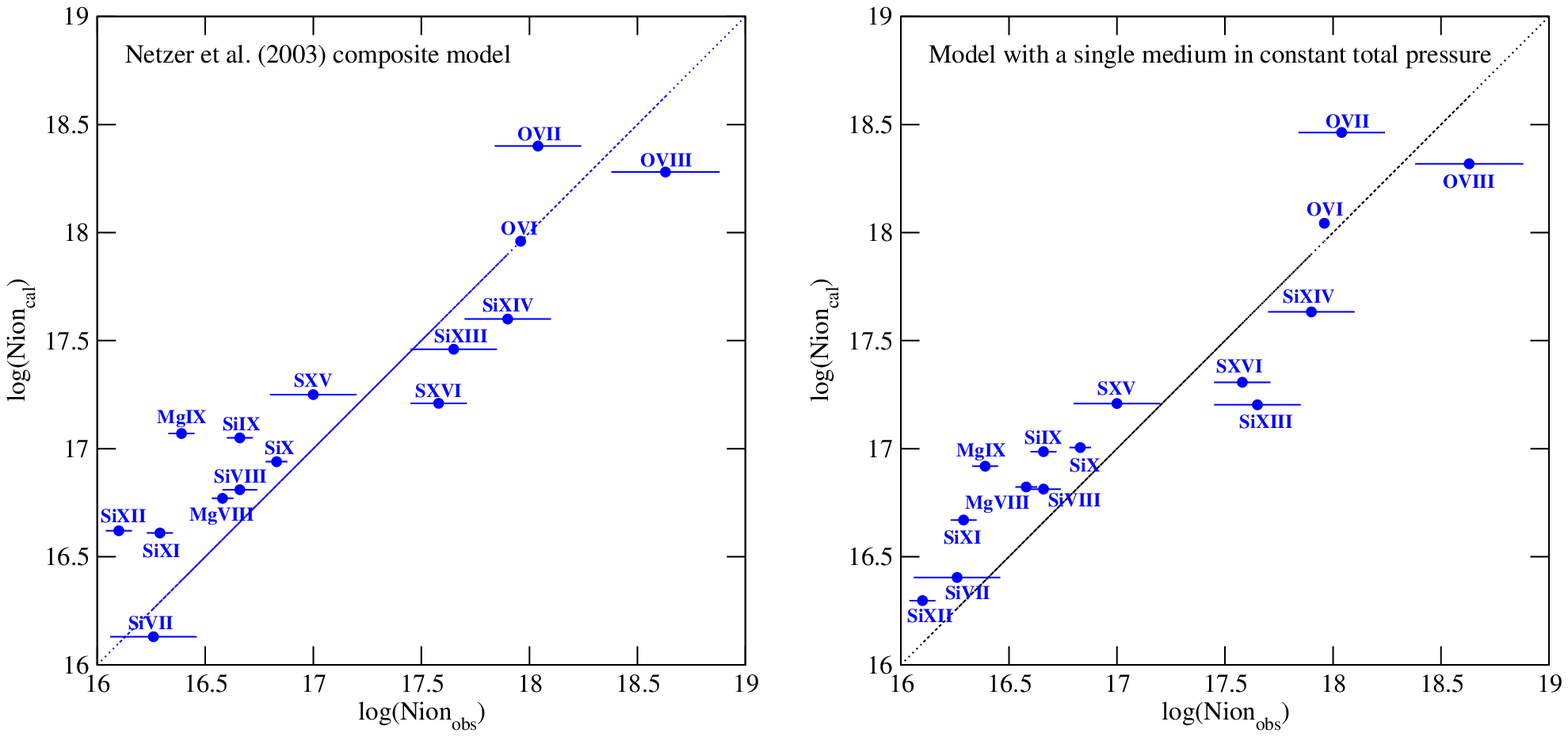}}
\hfill
\parbox[b]{45mm}{
\caption{Computed ionic column densities (in~\NHun) for the Netzer et 
al. (2003) composite constant density model (left panel) and for 
our constant total pressure model (right panel) \vs\ the 
``observed" column densities determined through a curve of 
growth analysis.  
The observed \ion{O}{vi} column density value corresponds to a 
lower limit, only. It was therefore replaced by the Netzer et al. 
computed value, which explains the absence of an error bar 
for this ion. The dotted line shows the diagonal on which the 
calculated values equal the observations.\vspace{2mm} 
} 
\label{ion}
}
\end{flushleft}
\end{figure*}

Our results can be compared to those of Netzer et al. (2003), 
who provide a detailed 
modelling of the WA in NGC~3783; these authors found three 
constant density components with different $\xiup$ and 
\NH\ values. 
A full comparison between the 
multi-component model of Netzer et al. (as computed by TITAN)  
and our constant total pressure model will also be presented 
in a forthcoming paper.  
Figure~\ref{ion} displays an example with the computed 
ionic column densities deduced from our best model and 
from the composite model of Netzer et al. 
\vs\ the ``observed'' ones determined through the  
curve of growth analysis in Netzer et al. (2003). The 
figure illustrates the good agreement between a single 
medium in constant total pressure and the three-zone  
model in constant density showing that, within the 
uncertainties, both models should lead to  similar absorption 
spectra; this comparison used the same incident SED and abundances. 
Although Fig.~\ref{ion} only shows 15 ions (for comparision with 
Netzer et al.), 
our model provides ionic column densitites for all the other ions in 
the atomic data, which include the lower ionization species that  
give rise to absorption lines in the UV.

Based on our best model results, and on the object's 
bolometric luminosity ($L$\,$\sim$\,2\,10$^{44}$\,\ergs) and 
black hole mass ($M_{\rm BH}$\,$\sim$\,3\,10$^{7}$\,M$_{\sun}$)
(Peterson et al. 2004), we have calculated some quantities 
related to the WA.  
Assuming $n_{\rm H}$\,=\,10$^{5}$\,cm$^{-3}$, the sum of 
the thickness of all the clumps in the line-of-sight amounts to 
$\Delta R_{\rm cl}$\,$\sim$\,2\,10$^{17}$\,cm 
(a higher $n_{\rm H}$ would imply a smaller 
$\Delta R_{\rm cl}$). Knowing $\Delta R_{\rm cl}$, 
and to keep $\dot{M}_{\rm out}/\dot{M}_{\rm Edd}$\,$\le$\,1 
(where $\dot{M}_{\rm out}$ and $\dot{M}_{\rm Edd}$ are the outflow 
and Eddington mass accretion rates), the volumic factor 
$f_{\rm vol}$ should be $\le$\,0.5 (see Eq.~4 in Chevallier et al. 2006). 
As $f_{\rm vol}$ corresponds to the size ratio of the ``clumpy'' to 
``whole WA'' media ($\Delta R_{\rm cl}/\Delta R_{\rm WA}$), 
we can estimate $\Delta R_{\rm WA}$ as being $\ge$\,4\,$10^{17}$\,cm. 
Based on the previously given definition of $\xiup$, we 
calculate the distance to the WA to be  $R$\,$\leq$\,10$^{18}$\,cm 
(for $n_{\rm H}$\,=\,10$^{5}$\,cm$^{-3}$).
 
Given these values, the replenishment timescale ($R$/\vflow) 
in NGC~3783 is of the order of 5\,10$^2$ yrs, and the 
dynamical timescale,  several times higher.  
A satisfactory WA geometry could be obtained with 
$f_{\rm vol}$\,$\sim$\,0.2 and $n \ge$\,\vturb/\vtherm\,$\sim$\/10. 
The geometrical thickness of a single clump should thus 
be $\le$\,2\,10$^{16}$\,cm, resulting in a sound crossing 
time of only $\sim$\,50\,yrs; if  $n  >$\,10, this 
time will be reduced in proportion.  In such a  picture, 
the replenishment and dynamical times are larger than 
the sound crossing time in each clump. 
In consequence, the clumps can adjust promptly to any 
change in the pressure equilibrium of the whole system. 
We note that these timescales are still much longer than the 
variation time of the illuminating flux; it could thus 
be objected that the gas in NGC~3783 cannot reach a 
state where it is in total pressure equilibrium with the 
illuminating source. We suggest that, unless the flux variations 
are very large, the medium would adopt a ``quasi-pressure 
equilibrium" corresponding to an incident flux averaged over a long 
time. The flux and spectral variations should thus induce 
changes in the temperature and ionization  
equilibrium,  but the density structure would stay about 
the same. The way the spectrum would be modified 
will be discussed in a forthcoming paper.

\vspace{-2mm}
\section{Conclusions}

\begin{enumerate}
\item We have shown that the WA in NGC 3783 can be modelled by a 
single medium in total pressure equilibrium.

\item Our grid of models has provided a best result for 
$\xiup$\,=\,2500 $\rm{erg\,cm\,s^{-1}}$, 
$N_{\rm H}$\,=\,4\,10$^{22}$ $\rm{cm}^{-2}$, and cosmic abundances. 


\item Our work opens new prospects for the future use 
of the TITAN code by the community through larger grids of 
constant total pressure models to be made available via XSPEC 
and/or via Virtual Observatory facilities.

\end{enumerate}

\begin{acknowledgements}
We acknowledge grant BPD/11641/2002 of the FCT, Portugal;  
grant 2P03D00322 of the PSCSR, Poland;  support from  
LEA and astro-PF, Poland-France, and HBS, Germany. 
\end{acknowledgements}


\begin{thebibliography}{}

\bibitem []{} Allen, C. W. 1973, Astrophysical quantities. 
U. London, Athlone Press





\bibitem{} Behar, E., Rasmussen, A. P., Blustin, A.~J., et al. \   
2003,  ApJ, 598, 232 

\bibitem[Blustin(2004)]{2004PhDT........24B} Blustin, A.~J. \ 2004, 
Ph.D.~Thesis, 
University College London, UK

\bibitem{} Chelouche, D., \& Netzer, H. \ 2005, ApJ, 625, 95

\bibitem{} Chevallier, L., Collin, S., Dumont, A.-M., et al. \ 2006, A\&A, 
accepted (astro-ph/0510700) 


\bibitem{}Collin, S., Dumont, A.-M., \&  Godet, O. \ 2004, A\&A, 419, 877

\bibitem[Dumont et al.(2000)]{2000A&A...357..823D} 
Dumont,\,A.-M.,\,Abrassart,\,A.,\,\&\,Collin,\,S.\,2000,\,\aap,\,357,\,823 

\bibitem[Dumont et al.(2002)]{2002A&A...387...63D} Dumont, A.-M., Czerny, 
B., Collin, S., \& Zycki, P.~T. \ 2002, \aap, 387, 63 


 

\bibitem{}\nom, A. C., et al., in preparation

\bibitem[Grupe et al.(2003)]{2003AJ....126.1159G} Grupe,  D., Mathur, S., 
\& Elvis, M. \ 2003, \aj, 126, 1159 
 
\bibitem[Guainazzi et al.(2001)]{2001MNRAS.323...75G} Guainazzi, M., 
Marshall, W., \& Parmar, A.~N.  2001, \mnras, 323, 75 
 
\bibitem[Halpern 1984]{Halpern:1984}Halpern, J. P. \ 1984, ApJ, 281, 90

\bibitem[Kaspi et al.(2000)]{2000ApJ...535L..17K} Kaspi, S., Brandt, W.\,N., 
Netzer,\,H., et al. \
2000, \apjl, 535, L17 

\bibitem{}Kaspi, S.,  Brandt, W. N., Netzer, H., et al. \ 2001, ApJ, 554, 216

\bibitem{} Kaspi, S.,  Brandt, W. N., George, I. M., et al. \ 2002, ApJ,
  574, 643

\bibitem{} Kinkhabwala, A., Sako, M., Behar, E., et al. 2002, \apj, 575, 732 

\bibitem[]{} Ko, Y.-K., \& Kallman, T. R. \ 1994, ApJ, 431, 273

\bibitem[Krolik \& Kriss(2001)]{2001ApJ...561..684K} Krolik, J.~H., \& 
Kriss, G.~A.\ 2001, \apj, 561, 684 

\bibitem[]{} Krolik, J. H., McKee, C. F., \& Tarter, C. B. \ 1981, 
ApJ, 249, 422

\bibitem{} Krongold,\,Y., Nicastro,\,F., Brickhouse,\,N.\,S., et al.  
2003,\,ApJ,\,597,\,832


\bibitem[Matsumoto et al.(2004)]{2004ApJ...603..456M} 
Matsumoto,\,C., Leighly,\,K.\,M., \& Marshall,\,H.\,L.\,2004,\,\apj,\,603,\,456 


\bibitem{} Netzer, H.,  Kaspi, S., Behar, E., et al. \ 2003, ApJ, 599, 
  933 

\bibitem[Peterson et al.(2004)]{2004ApJ...613..682P} Peterson, B.\,M., 
  Ferrarese, L., Gilbert, K. M., et al. 2004, \apj, 613, 682 




\bibitem[]{} R\' o\. za\' nska, A., Czerny, B., Dumont, A.-M., \&
  Collin, S. \ 2002, MNRAS, 332, 799
 
\bibitem{} R\'o\.za\'nska, A., Goosmann. R., Dumont. A.-M., \& Czerny,
  B. \ 2006, A\&A, accepted (astro-ph/0512310)

\bibitem[Sambruna et al.(1997)]{1997ApJ...483..774S} 
Sambruna,\,R.\,M.,\,George,\,I.\,M.,\,Madejski,\,G.,\,et\,al.\,1997,\,\apj,\,483,\,774  

\end{thebibliography}
\end{document}